\documentclass{article}

\usepackage{amssymb}
\usepackage{amsmath}
\usepackage[cp1251]{inputenc}
\usepackage[dvips]{graphicx}
\usepackage{longtable}

\begin{document}

\title{\textbf{Improved efficiency of planar light converters based on Al$_{2}$O$_{3}$-YAG:Ce eutectic crystals: Physical Trends and Limits }}

\author{\textbf{S.V. Naydenov${}^{1}$, O.M. Vovk${}^{1}$, Yu.V. Siryk${}^{1}$, S.V. Nizhankovskyi${}^{1}$, I.M. Pritula${}^{1}$}}

\maketitle

\begin{center}
$^{1}$\textit{Institute for Single Crystals of National Academy of Sciences of Ukraine, \\ 60 Nauky Avenue, 61072 Kharkiv, Ukraine}
\end{center}

\begin{abstract}
A theoretical model of a crystalline converter of white light with internal scattering medium is proposed.  Obtained are the expressions for estimation of the light converter efficiency depending on its optical parameters (refractive index and coefficients of absorption at the wavelengths of the incident and re-emitted light), as well as on the geometrical dimensions and the scattering indicatrix of the optical system. The physical limits of improvement of the converter efficiency due to strong internal scattering, are established. Realized are the theoretical and experimental estimations of the efficiency of the light converter based on the crystals of Al${}_{2}$O${}_{3}$-YAG:Ce eutectic alloy which amounts to 16\%. This is twice as high in comparison with the efficiency of the light converter based on YAG:Ce single crystal.


\end{abstract}

\section{Introduction}

One of the promising present-day methods for obtaining of white light is the use of light converters which transform monochromatic light of blue LED (main emission of InGaN chip at $\lambda$ \~{}460~nm), or high-power laser diode (LD) into a mixture of the light partially transmitted at the same wavelength and the light converted in the yellow-green or in another spectral range [1]. The converter efficiency is defined by the total output light flow. The ratio of output fluxes of the spectral components influences the quality of white light (color temperature, color rendering coefficient, etc.). The materials conventionally used in light converters include scintillation glasses [2, 3], transparent ceramics [4, 5], or phosphors [6, 7]. One of the main factors that defines low light converter efficiency is the effect of total internal reflection which transforms a light converter of regular shape into a wave guide that captures a major portion of light rays inside itself. This effect can be suppressed by different methods, such as profiling of the emitting surface of light converter [8, 9], introduction of scattering metal-oxides particles of micron size into the converter material [10], creation of internal strongly scattering medium due to texturing of the material, in particular, the use of an eutectic compound consisting of several phase components [11-14]. For the ceramic composites Al${}_{2}$O${}_{3}$-YAG:Ce and the eutectic Al${}_{2}$O${}_{3}$-YAG:Ce crystals there was revealed an essential improvement of their lighting parameters [15, 16].

Proposed in the present work is the physical model of the light converter with internal light scattering. The theoretical expressions for the determination of the energy efficiency of the light converter are obtained, and the physical limit of improvement of this parameter is estimated. For the light converters based on Al${}_{2}$O${}_{3}$-YAG:Ce eutectics grown by the method of horizontally directed  crystallization (HDC), there are measured some lighting parameters. The comparison of  the converter efficiency and the quality of white light, as well as the performed theoretical estimations, testify that such a eutectic is promising for the development of new material for white light converters.

\noindent

\section{Theoretical model of light converter with internal scattering }

Consider the physical model of light converter which has the geometry of plane-parallel plate with the thickness $H$ and the area $S$. Light conversion is realized in transmittance geometry when the incident light (usually the one of blue light emitting diode at $\lambda $=460~nm) normally falls on one of the converter sides. Inside the converter, the light is partially absorbed by the converter material and re-emitted at another wavelength (e.g. for YAG:Ce crystals at  $\lambda $=540 nm (of Ce${}^{3+}$ion emission). We claim that the distribution of scintillations caused by the conversion is uniform and isotropic. The former statement is valid for those light converters that thickness is less than the length $l_{b} $ of bulk absorption of incident radiation, i.e. for the samples with a thickness of several millimeteres. The statement concerning isotropy of the angular distribution of scintillations is generally accepted. Let us dwell on the simplest scheme of the conversion which considers only the emission at two wavelengths: at the maxima of absorption $\lambda _{b} $ and luminescence (conversion) $\lambda _{c} $ of the light converter. Taking into account the spectral dependence of light intensity complicates the considered problem. But it is not a fundamental restriction, since a real emission spectrum can be always expanded in monochromatic waves with subsequent analysis of each of them based on the estimations analogous to those obtained below.

The output light intensity of the converter is expressed as
\begin{equation} \label{1)}
I=I_{b} +I_{c}  ,
\end{equation}
where $I_{b} $ and $I_{A} $ are the intensities of light at the wavelengths $\lambda _{b} $ and $\lambda _{A} $, respectively. Assume that in the general case the converter material is an internal scattering medium. Now we will not specify the nature of the scattering centers and the character of the scattering indicatrix. When propagating inside the converter, the light beams may undergo scattering. In the case of the presence of a dispersed medium consisting of reflecting micro-particles, the scattering is analogous to the Rayleigh scattering. In the case of eutectic crystals, this is either the scattering  on optical inhomogeneities (always present in such systems) or the micro-scattering bound up with transmission of a multi-phase medium which consists of several (two or more) phases with different values of bulk refractive index. At multiple transfers from one phase to another, the light beams undergo ``internal refraction'' and may change their initial direction of motion. This is schematically shown in Fig. ~1. In contrast to a dispersed medium of point phase scatters, the phases of eutectic are located inside the crystal where they interlace in a complex manner and have a fractal border.

Thus, the intensity of transmitted light $I_{b} $ is the sum of two components
\begin{equation} \label{2)}
I_{b} =I_{b}^{\left(0\right)} +I_{b}^{\left(s\right)}
\end{equation}
where $I_{b}^{\left(0\right)} $and $I_{b}^{\left(s\right)} $ correspond to direct and scattered light, respectively, the latter being a result of multiple reflections from internal scatterers and external borders of the system. These intensities are bound up with the initial intensity $I_{0} $ of incident light by the following relations
\begin{equation} \label{3)}
I_{b}^{\left(0\right)} =\tau _{b}^{\left(0\right)} I_{0} ;\quad I_{b}^{\left(s\right)} =\tau _{b}^{\left(s\right)} I_{0}  ,
\end{equation}
where $\tau _{b}^{\left(0\right)} $ and $\tau _{b}^{\left(s\right)} $ are the coefficients of light output for direct and scattered light, respectively, at the wavelength $\lambda _{b} $. It is convenient to introduce the coefficient of total light output $\tau _{b} $ for incident light
\begin{equation} \label{ZEqnNum626475}
I_{b} =\tau _{b} I_{0}  ,
\end{equation}
where
\begin{equation} \label{5)}
\tau _{b} =\tau _{b}^{\left(0\right)} +\tau _{b}^{\left(s\right)}  .
\end{equation}

By analogy, for the light converted at $\lambda _{A} $
\begin{equation} \label{6)}
I_{A} =I_{A}^{\left(0\right)} +I_{A}^{\left(s\right)}  ,
\end{equation}
where
\begin{equation} \label{7)}
I_{A}^{\left(0\right)} =\tau _{A}^{\left(0\right)} I_{0}^{*} ;\quad I_{A}^{\left(s\right)} =\tau _{A}^{\left(s\right)} I_{0}^{*}
\end{equation}
Here  $I_{0}^{*} $ is the initial intensity of the light emitted at $\lambda _{A} $ in the converter during passage of incident light in it at $\lambda _{b} $. As follows from the light energy conservation law, $I_{0}^{*} $ is proportional to the portion of  incident light that has not left the converter, i.e. the value $\left(1-\tau _{b} \right)$, as well as to the conversion efficiency $k_{c} $. As a rule, during the making of the converter it is necessary to provide \~{}90-95\% efficiency of quantum luminesce of Ce${}^{3+}$ ions.

At a good approximation, in the considered case
\begin{equation} \label{8)}
I_{0}^{*} \approx \left(1-\tau _{b} \right)I_{0}  .
\end{equation}
Now introduce the total light collection coefficient $\tau _{A} $ for the converter light and obtain
\begin{equation} \label{9)}
I_{A} =\tau _{A} I_{0}^{*}  ,
\end{equation}
where
\begin{equation} \label{10)}
\tau _{c} =\tau _{c}^{\left(0\right)} +\tau _{c}^{\left(s\right)}  .
\end{equation}
In the equivalent form:
\begin{equation} \label{ZEqnNum494250}
I_{c} =\; \eta _{A} I_{0}  ,
\end{equation}
where  $\eta _{A} $  is the conversion(al) efficiency of the light converter
\begin{equation} \label{11)}
\eta _{A} =k_{c} \left(1-\tau _{b} \right)\tau _{A}  .
\end{equation}

The total light converter efficiency $\eta $ (expressed in relative units, i.e. as a percentage) is determined from the expression
\begin{equation} \label{ZEqnNum132139}
\eta \equiv \frac{I_{b} +I_{c} }{I_{0} } =\tau _{b} +\eta _{c}  .
\end{equation}
The ``color formula'' of the light converter, i.e. the ratio of the conditional ``blue'' $B$ and ``yellow'' Y ``color coordinates'' of the output light is the following:
\begin{equation} \label{ZEqnNum296912}
B:Y\equiv I_{b} :I_{c} =\tau _{b} :\eta _{c}  .
\end{equation}
In principle, relations \eqref{ZEqnNum132139} and \eqref{ZEqnNum296912} solve the problem of estimation of the efficiency and spectral composition  of the light emitted by the converter. As follows from \eqref{ZEqnNum296912}, for this purpose it is necessary to calculate the coefficients of total light collection $\tau _{b} $ and $\tau _{A} $(in the frame of an appropriate model), or to measure them.

From the viewpoint of mathematics, the calculations of $\tau _{b} $ and $\tau _{A} $ not much differ (except distinctions in the wavelength of the considered monochromatic light and in the initial conditions for incident and re-emitted light).

Therefore, it is sufficient to solve the problem of light collection only for one of the partial components of the light emitted by the convertor.

In the approximation of ideal conversion when $k_{c} \approx 1$ let us rewrite relation \eqref{ZEqnNum132139} in the form
\begin{equation} \label{ZEqnNum607039}
\eta \approx \tau _{b} +\left(1-\tau _{b} \right)\tau _{c} =\tau _{b} +\tau _{c} -\tau _{b} \tau _{c}  .
\end{equation}
As will be seen below, $\tau _{b} $ and $\tau _{A} $ are small values, since, as a rule, the light output for a medium with a strong total internal reflection (as in the considered case) is not higher than several tens of percent, i.e. the condition $\tau \ll 1$ is fulfilled well enough. Therefore, the nonlinear term $\tau _{b} \tau _{c} $ in eq. \eqref{ZEqnNum607039} is to be neglected:
\begin{equation} \label{16)}
\eta \approx \tau _{b} +\tau _{c} \quad {\rm if}\quad \tau \ll 1 .
\end{equation}
So, the efficiency $\eta $ of the transformation of light energy in the converter is defined by the total light collection at the wavelengths of the incident and converted light. By analogy, when taking into account the spectral dependence for the incident $I_{0} \left(\lambda \right)$ and output $I\left(\lambda \right)$ light for the converter efficiency, one may assume that
\begin{equation} \label{ZEqnNum992534}
\left\langle \eta \left(\lambda \right)\right\rangle \approx \tau _{b} +\left\langle \tau _{c} \left(\lambda \right)\right\rangle \quad {\rm if}\quad \left\langle \tau \left(\lambda \right)\right\rangle \ll 1 ,
\end{equation}
where $\left\langle \ldots \right\rangle =\int _{\Delta \lambda }\ldots d\lambda  $ denotes the spectral averaging over all the wavelengths; $\tau _{c} \left(\lambda \right)={I_{c} \left(\lambda \right)\mathord{\left/ {\vphantom {I_{c} \left(\lambda \right) I_{0} \left(\lambda \right)}} \right. \kern-\nulldelimiterspace} I_{0} \left(\lambda \right)} $ is the spectral value of the coefficient of light collection; $\Delta \lambda $, the spectral range of emission. The maximum light converter efficiency $\eta =\eta _{\max } $ is obviously at the maximum light collection  $\tau =\tau _{\max } $.

Determination of the light collection efficiency $\tau $ is a difficult task, since, as a rule, light collection depends on many parameters and conditions. This value can be found by direct measurements (using test light sources) on test samples with preset chemical composition and geometry with a subsequent generalization of the obtained results. Alternatively, $\tau $ can be calculated by the Monte Carlo method using different computer codes (see e.g. [17]). In some particular cases the coefficient of light collection can be calculated directly, or in the frames of the diffusion model (see e.g. [18]). However, in all these cases it is practically impossible to establish regularities in the light collection, especially in the presence of an internal highly/strongly scattering medium. Therefore, we choose the phenomenological approach which uses the model of chaotic light collection [19]. Within this approach, the coefficient of the light collection is calculated under the assumption of strong chaos of light flows [19-21]. The said property manifests itself in the absence of strong correlations between multiply reflected flows that emerge from the system after a different (finite) number of reflections.

The coefficients of direct light collection are expressed by the formulas:
\begin{equation} \label{ZEqnNum803382}
\tau _{b}^{\left(0\right)} =q_{b0} \left(1-R_{0b} \right)^{2} T_{0b}  ,
\end{equation}
\begin{equation} \label{ZEqnNum334784}
\tau _{c}^{\left(0\right)} =q_{c0} \left(1-R_{0c} \right)T_{0c}  ,
\end{equation}
where  $q_{0} $ is the aperture (probability) of direct output of light beams without direct reflections; $R_{0b,c} $, the reflection coefficient for the Fresnel losses at  (normal)  crossing of each external light converter border
\begin{equation} \label{20)}
R_{0b,c} \approx \left[\frac{n\left(\lambda _{b,c} \right)-1}{n\left(\lambda _{b,c} \right)+1} \right]^{2}  ,
\end{equation}
Here $n=n\left(\lambda \right)$ is the relative refractive index; $T_{0b,c} $, the coefficients of (bulk) transmission for direct and converted light in optical medium in the absence of scatterers:
\begin{equation} \label{ZEqnNum602250}
T_{0b} \approx \exp \left(-\alpha _{b} H\right),\quad T_{0c} \approx \exp \left(-\alpha _{c} {H\mathord{\left/ {\vphantom {H 2}} \right. \kern-\nulldelimiterspace} 2} \right) ,
\end{equation}
where  $\alpha _{b} $ and $\alpha _{A} $ are the coefficients of linear light absorption at the wavelengths $\lambda _{b} $ and $\lambda _{c} $, respectively; $H$, the converter thickness.  The distinction between these expressions  is bound up  with the fact that in the former  incident light passes through the converter entirely, whereas in the latter it is emitted inside the converter at different depths (thereat, the effective depth of light passage is equal to ${H\mathord{\left/ {\vphantom {H 2}} \right. \kern-\nulldelimiterspace} 2} $). The apertures of direct light output $q_{b0} ,q_{c0} $ are defined by the size/dimensions of the system, the incident beam geometry, the cone of total internal reflection and the existence of ``light guiding'' channels between the scatterers.

Summation of light flows for multiply scattered light (similarly to the method [19]) leads to the formulas:
\begin{equation} \label{ZEqnNum333312}
\tau _{b}^{\left(s\right)} =\left(1-q_{b0} \right)\left(1-R_{0b} \right)\frac{q_{b} \left(1-R_{b} \right)\tilde{T}_{b} }{1-\left(1-q_{b} \right)R_{b} \tilde{T}_{b} }  ,
\end{equation}
\begin{equation} \label{23)}
\tau _{c}^{\left(s\right)} =\left(1-q_{c0} \right)\; \frac{q_{c} \left(1-R_{c} \right)\tilde{T}_{c} }{1-\left(1-q_{c} \right)R_{c} \tilde{T}_{c} }  ,
\end{equation}
where $q_{b,c} $ is the average aperture of the output of multiply scattered light (the probability that the beam will reach the converter output window); $\tilde{T}_{b,c} $ is the transparency of the system taking into account the internal scattering:
\begin{equation} \label{ZEqnNum790731}
\tilde{T}_{b} =\exp \left(-\alpha _{b} L_{s} \right);\quad \tilde{T}_{c} =\exp \left(-\alpha _{c} L_{s} \right) ,
\end{equation}
where $L_{s} $is the transport length, i.e. the average free pass length of a light beam in a scattering medium before it hits one of external borders. The length $L_{s} $ is bound up with the length $l_{s} $ of the beam free pass/run between two sequential reflections from the internal scatterers/diffusers by the relation
\begin{equation} \label{25)}
L_{s} =\gamma _{s} l_{s}  ,
\end{equation}
where $\gamma _{s} $ is the scattering factor which can be considered a phenomenological parameter. The coefficient $R_{b,c} =\left\langle R\left(n\left(\lambda _{b,c} \right),\theta \right)\right\rangle $, in contrast to $R_{0b,c} $, is the average value of the true coefficient of reflection when a light beam hits the border at different angles $\theta $;
\begin{equation} \label{26)}
\begin{array}{l} {R\left(n,\theta \right)=\left[\frac{n\cos \theta -\sqrt{1-n^{2} \sin ^{2} \theta } }{n\cos \theta +\sqrt{1-n^{2} \sin ^{2} \theta } } \right]^{2} =} \\ {=\left(\frac{n^{2} +1}{n^{2} -1} \right)^{2} \left[1-\frac{2n}{n^{2} +1} \cos \theta \sqrt{1-n^{2} \sin ^{2} \theta } -\frac{2n^{2} }{n^{2} +1} \sin ^{2} \theta \right]^{2} } \end{array}
\end{equation}
in the absence of total internal reflection
\begin{equation} \label{27)}
\left|\sin \theta \right|\le \sin \theta _{c} =\frac{1}{n}  ,
\end{equation}
Thereat, in the opposite case ${1\mathord{\left/ {\vphantom {1 n}} \right. \kern-\nulldelimiterspace} n} \le \sin \theta \le 1$ we have $R=1$ by definition. Note that $R+\tilde{T}\ne 1$, since these values describe different processes: reflection at the border and transmission in the bulk of the system.

For simplification, let us neglect possible spectral dependence of the apertures of the scattered light, i.e. $q_{b} \approx q_{c} =q$, as well as the distinction of the average values of the reflection coefficient from their values at the normal incidence, i.e. $R_{0b} \approx R_{b} $ and $R_{0c} \approx R_{c} $. Then
\begin{equation} \label{ZEqnNum244054}
\tau _{b} =q_{0b} \left(1-R_{b} \right)^{2} T_{0b} +\left(1-q_{0b} \right)\left(1-R_{b} \right)^{2} \frac{q\tilde{T}_{b} }{1-\left(1-q\right)R_{b} \tilde{T}_{b} }  ,
\end{equation}
\begin{equation} \label{29)}
\tau _{A} \approx q_{0c} \left(1-R_{c} \right)T_{0c} +\left(1-q_{0c} \right)\left(1-R_{c} \right)\; \frac{q\tilde{T}_{c} }{1-\left(1-q\right)R_{c} \tilde{T}_{c} }  .
\end{equation}
As a rule, for the converter with a high refractive index $n\ge 1.8$ Fresnel reflection losses are low: $R_{b,c} \ll 1$. In this approximation, as well as taking into account the additional conditions $T_{0b} \approx \tilde{T}_{b}^{2} $ and $T_{0A} \approx \tilde{T}_{A} $ (valid at $\gamma _{s} \sim 1$), we obtain:
\begin{equation} \label{ZEqnNum103204}
\tau _{b} =q_{0b} \tilde{T}_{b}^{2} +\left(1-q_{0b} \right)q\tilde{T}_{b}  ,
\end{equation}
\begin{equation} \label{ZEqnNum367747}
\tau _{A} \approx q_{0c} \tilde{T}_{c} +\left(1-q_{0c} \right)\; q\tilde{T}_{c}  .
\end{equation}

In the main, the aperture  $q_{0c} $ is defined by the cone of total internal reflection  $q_{0c} \sim f_{0} $, where $f_{0} $ is the angular aperture of the direct output of the re-emitted light. In the considered planar geometry
\begin{equation} \label{ZEqnNum897741}
f_{0} =f_{F} \equiv 1-\sqrt{1-\frac{1}{n^{2} } }  ,
\end{equation}
that corresponds to the Fresnel (mirror) indicatrix of scattering at the border. Note that $f_{F} \sim {1\mathord{\left/ {\vphantom {1 2n^{2} \ll 1}} \right. \kern-\nulldelimiterspace} 2n^{2} \ll 1} $ at sufficiently high values of the reflective index $n>1.8$.

The aperture of light output with multiple light scattering $q=g\, f$ is the product of the geometrical $g$ and angular $f$apertures. By analogy with the data [19], the latter is determined from the expression
\begin{equation} \label{ZEqnNum667887}
f_{D} =\sum _{k=0}^{+\infty }f_{k} \left\{1-\left[1-n^{-2} \right]^{\, {\left(k+1\right)\mathord{\left/ {\vphantom {\left(k+1\right) 2}} \right. \kern-\nulldelimiterspace} 2} } \right\}  ,
\end{equation}
 where $f_{k} $ are the amplitudes of expansion of (phenomenological) angular scattering indicatrix $f\left(\theta \right)$presented in the form of partial expansion in partial harmonics
\begin{equation} \label{34)}
f_{D} \left(\theta \right)=\left(2\pi \right)^{-1} \sum _{k=0}^{+\infty }\left(k+1\right)f_{k} \cos ^{k} \theta   ,
\end{equation}
with the total/full normalization $\sum _{k}f_{k}  =1$; $\theta $ is the polar angle in the plane perpendicular to the light converter plane. For the limiting case of the Lambertian (equally bright in all directions) scattering amplitude $f_{L} \left(\theta \right)\propto \cos \theta $ we obtain
\begin{equation} \label{ZEqnNum650819}
f_{L} =\frac{1}{n^{2} }  .
\end{equation}
The geometrical aperture $g$ (the probability of light output in configuration space) is expressed as
\begin{equation} \label{36)}
g=c_{s} \left(\frac{S_{out} }{S} \right) ,
\end{equation}
where $S_{out} $ is the output window area; $S$ is the total surface of the detector area; $c_{s} $ is the scattering factor (geometrical shadow), due to which the direction of ray propagation in the converter is not straight. As a rule, for the estimations it is assumed that $c_{s} \sim 1$. For a planar light converter shaped as a thin plate $H\ll L$ ($L$is the length) that works in the geometry of transmitted light at $S_{out} \approx {S\mathord{\left/ {\vphantom {S 2}} \right. \kern-\nulldelimiterspace} 2} $, we obtain $g\approx {1\mathord{\left/ {\vphantom {1 2}} \right. \kern-\nulldelimiterspace} 2} $. The converter may also have another geometry (e.g. the one of photometric sphere or hemisphere) when it is almost the whole of its surface that emits. In this case $g\approx 1$.

\section{Physical limits of light converter efficiency }

As follows from eqs. \eqref{ZEqnNum103204}-\eqref{ZEqnNum367747}, $\tau _{b,c} $ are monotonic functions of the parameters  $q_{0} $ and $q$. In the absence of internal scattering their growth is restricted by the limiting values $q_{0b} =1$ and $q_{0c} \approx q_{c} =gf_{F} $. In the presence of scattering $q_{0b} \approx 1$, too. However, the factor of multiple scattering  $q_{0c} \approx q_{c} =gf_{D} $ may be essentially higher and reach its maximum value  $gf_{L} $ at a strong chaotic ray scattering. As a result, after internal scattering the beam that has initially propagated within the cone of total internal reflection, reaches the converter border at such an angle when it can emerge from the converter.

In view of eqs. \eqref{ZEqnNum103204}-\eqref{ZEqnNum367747}, the efficiency of the converter without scattering, i.e. at $q_{0b} =1,\; q_{0b} \approx q_{0c} =gf_{F} ,\; q_{b} =q_{c} =0$ is equal to
\begin{equation} \label{ZEqnNum698957}
\eta _{F} \approx \left(T_{0b} +gf_{F} \tilde{T}_{c} \right) ,
\end{equation}
whereas at a strong scattering when $q_{0b} =1,\; q_{0b} \approx q_{0c} =gf_{F} ,\; q_{b} \approx q_{c} =gf_{D} $ it will be expressed as
\begin{equation} \label{38)}
\eta _{D} \approx \eta _{F} +g\left(1-gf_{F} \right)f_{D} \left(\tilde{T}_{b} +\tilde{T}_{c} \right) .
\end{equation}

Assume that for the converter in which the scattering is caused by the presence of a dispersed medium emerging from internal opaque scatterers (such as ceramics containing agglomerates of spherical metallic particles, see e.g. [10]), or by small-angle scattering on microscopic inhomogeneities of heterogeneous phase composition, as it takes place in eutectic alloys, at the condition of strong scattering the transparency at the absorption wavelength is minimal: $\tilde{T}_{b} \ll 1$, whereas at the re-emission wavelength it is maximal: $\tilde{T}_{c} \approx 1$. For instance, for YAG:Ce-based light converter with a small thickness $H\sim 10^{-1} cm$ and typical absorption coefficients $\alpha _{b} \sim 10^{1} \, cm^{-1} $ and $\alpha _{A} \sim 10^{-2} \, cm^{-1} $, this condition is certainly fulfilled. In this case
\begin{equation} \label{39)}
\eta _{F} \approx gf_{F} ;\quad \eta _{D} \approx \eta _{F} +g\left(1-gf_{F} \right)f_{D}  .
\end{equation}

Enhancement factor $G$ of the converter light output at the change-over from a conventional transparent crystal to a strongly scattering/dispersive medium is equal to
\begin{equation} \label{ZEqnNum358569}
G\equiv \frac{\eta _{D} }{\eta _{F} } =1+\frac{g\left(1-gf_{F} \right)f_{D} }{\eta _{F} }  .
\end{equation}
The limiting value of the parameter $G=G_{*} $ is achieved in the case of the ideal (Lambert) scattering $f_{D} \approx f_{L} $:
\begin{equation} \label{ZEqnNum369161}
G_{*} =G_{*} \left(g,n\right)=1+\frac{\left(1-gf_{F} \right)f_{L} }{f_{F}^{2} }  .
\end{equation}
When substituting expressions  \eqref{ZEqnNum897741} and \eqref{ZEqnNum650819} into \eqref{ZEqnNum369161} we obtain
\begin{equation} \label{42)}
G_{*} \left(g,n\right)=1+\frac{1}{n^{2} } \left[n\left(1-g\right)+g\sqrt{n^{2} -1} \right]\left[n+\sqrt{n^{2} -1} \right] .
\end{equation}
Gain in the efficiency of the energy transformation by the scattering medium is the more noticeable, the higher its relative refractive index $n$, i.e. the more the effect of total internal reflection is compensated, and the smaller the output window area and, consequently, the smaller the parameter $g$ . For the ideal medium $n=1$ and the light collection over the whole of the converter surface ($g=1$) such a gain is absent: $G_{*} =1$. For higher  $n\gg 1$  $G_{*} \to 3$, i.e. for any material the maximum gain at enhancement of the converter efficiency does not exceed 300\%.

Presented in Fig. 2 are the typical theoretical dependencies of the efficiency for the a converter based on a conventional transparent medium (a single crystal or a transparent ceramics), and for a medium with strong internal scattering (a eutectic alloy, or a composite with dispersing medium consisting of micro-scatterers/diffusers). We have chosen a flat geometry of the converter with energy transformation in transmitted light and the aperture $g=0.5$. Enhancement of the light output for a weakly refracting medium $n\sim 1$ is caused by the fact that strong scattering not only favors release of the captured light (suppression of the effect of total internal reflection) but also essentially reduces the losses by the back reflection of the re-emitted light in the direction opposite to the one of the initial light incident on the converter.

For the light converter based on YAG:Ce crystal and Al${}_{2}$O${}_{3}$-YAG:Ce eutectic with the thickness $H=0.1\, cm$, the refractive index $n=1.83$ (with the dispersion neglected), the absorption coefficients  $\alpha _{b} =50\, cm^{-1} $ and $\alpha _{c} =0.01\, cm^{-1} $, expressions  \eqref{ZEqnNum698957}-\eqref{ZEqnNum358569} give the theoretical estimations $\eta _{F} \approx 8.8\% $ and $\eta _{D} \approx 22.5\% $, i.e. the theoretical gain of light output  may be $G=256\% $. As found in the experiment (see below) for a transparent crystal $\eta _{F} \sim 8\% $, that is close to the theoretical estimation. Thereat for an eutectic crystal (at strong scattering) the value of light output is much lower: $\eta _{D} \sim 16\% $, i.e. the energy gain is \~{}200\%, since for eutectic light transmission and light scattering are not ideal.

\section{Lighting parameters of  Al${}_{2}$O${}_{3}$-YAG:Ce eutectics }

Transparent YAG:Ce single crystal and  Al${}_{2}$O${}_{3}$-YAG:Ce eutectic were grown by the method of HDC [22]. During crystallization of the eutectic the pulling rate was changed. The grown ingots were used to obtain the following samples: Sample \#1 -- YAG:Ce single crystal with polished surfaces, Sample \#2 -- YAG:Ce single crystal with rough (matted) surfaces, Samples \#3-5 -- Al${}_{2}$O${}_{3}$-YAG:Ce eutectic grown at different crystallization rates. All the grown crystals had the dimensions 7x7x1 mm$^{3}$.

Fig.~3 presents the typical photographs and scanning electron microscopic (SEM)images of the fragments of Al${}_{2}$O${}_{3}$-YAG:Ce sample. A distinctive feature of the eutectic morphology is a pronounced dispersed structure (with possible additional textures and inhomogeneities), that consists of the phases YAG:Ce and Al${}_{2}$O${}_{3}$ mutually penetrating one another. The former of them converts and transmits light, the latter plays the role of intermediate light scattering medium (the refractive index of sapphire Al${}_{2}$O${}_{3}$ : $n=1.77$ is a little lower than the one of  YAG : $n=1.83$).

The lighting characteristics of the samples were determined using a lighting stand HAAS-2000(EVERFINE). The power of light flow from the source Blue-LED was 2.35~mW with the emission wavelength maximum $\lambda $=460 nm and FWHM=30 nm. The data on the lighting measurements of the efficiency of the converters based on YAG:Ce single crystals and Al${}_{2}$O${}_{3}$-YAG:Ce eutectics are presented in Table.

The obtained results testify to the advantage of the light converters based on the eutectic over those based on the conventional transparent YAG:Ce crystal. Here the absolute efficiency is defined as the ratio of the energy flows of the output and incident light. It corresponds to the earlier introduced theoretical parameter $\eta $. As is seen, for polished YAG:Ce crystals (Sample 1) the efficiency is consistent with the theoretical estimation. For the eutectic crystals the efficiency (\~{} 16\%) is lower than the theoretical one (\~{} 22\%), but it is twice as high as the efficiency of the polished crystals. Moreover, the eutectic samples have a better color temperature, and their color coordinates are closer to white color (the sum of the color coordinates is closer to 1, $x+y\approx 1$). High efficiency is also characteristic of the crystals with matted output surface (Sample 2). For such crystals, the light output increases and the spectral characteristics are improving as the surface roughness grows and the scattering indicatrix becomes closer to the Lambert indicatrix.[23], [24]. This is bound up with suppression of the effect of total internal reflection on the converter surface.

\section{Conclusions}

The proposed theoretical model for  description of the efficiency of the light converters based on crystalline materials confirms that the said characteristic considerably rises in the presence of  strong internal scattering  of light flows. Such a scattering medium may be: 1) mattered light emitting surface with strong surface light scattering; 2) internal dispersed medium which consists of micro-particles of metals or metal oxides reflecting light; 3) multi-phase medium of eutectic alloy with optical characteristics similar to those of typical turbid  medium. The parameters of the light converter based on the eutectic Al${}_{2}$O${}_{3}$-YAG:Ce  determined theoretically and experimentally are in qualitative conformity. Thereat, the efficiency of the converter based on the eutectic is at least twice as high in comparison with the corresponding characteristic of the converter based on transparent YAG:Ce crystal. Optimization of the morphology of Al${}_{2}$O${}_{3}$-YAG:Ce eutectic may lead to further increase of energy (absolute) efficiency of the converter up to its theoretical limit ( \~{}269\%) as against the one of YAG:Ce --based converter. Its limit approaches for light converter having refraction coefficient $n=1.83$ and emitting ahead by 2$\pi $-sphere (half of all surface of light converter). Moreover, it seems promising to dope or to co-dope the eutectics with ions of rare-earth and transient metals (e.g. gadolinium) for shifting the spectral emission band from the yellow-green to the red-green region of the visible spectrum.  The latter provides a better agreement of the emission spectrum of the converter with the curve of visibility, and may lead to additional increase of the index of color rendering and light efficiency of the converter.

Finally, the proposed theoretical approach and experimental methods for the obtaining of new crystalline media for production of white light may be applied for development of new light converters not only in the optical scheme with transmitted LED light, but also in the scheme with reflected light, i.e. for creation of high-power WLED on the base of high-power LD. This will be the subject of our further research.

\section{References}

\begin{enumerate}
\item  L. Chen, C.-C. Lin, C.-W. Yeh, R.-S. Liu, Materials, 3, 2172 (2010). DOI:10.3390/ma3032172/

\item  X. Ma, X. Li, J. Li, et al., Nature Communications, 9, 1175 (2018). DOI:10.1038/s41467-018-03467-7/

\item  Y. Yuan, D. Wang, B. Zhou, et al., Optical Materials Express, 8 (9), 2760 (2018). DOI:10.1364/OME.8.002760/

\item  Q.-Q. Zhu, S. Li, Q. Yuan, et al., Journal of the European Ceramic Society, 41, 735 (2021). DOI:10.1016/j.jeurceramsoc.2020.09.006/

\item  Q.-Q. Zhu, Y. Meng, H. Zhang, et al., ACS Appl. Electron. Mater., 2 (8), 2644 (2020). DOI:10.1021/acsaelm.0c00512/

\item  A. Revaux, G. Dantelle, placeS. Brinkley, et al., Proc. of SPIE, 8102, 81020R-1 (2011). DOI:10.1117/12.892717/

\item  Y.R. Tang, S.M. Zhou, X.Z. Yi, et al., Optics Letters, 40 (23), 5479 (2015). DOI:10.1364/OL.40.005479/

\item  G. Singh, D.S. Mehta, Journal of Information Display, 15 (2), 91 (2014). DOI:10.1080/15980316.2014.903211/

\item  D. Huh, W. Kim, K. Kim, et al., Nanotechnology, 31 (14), 144003 (2019). DOI:10.1088/1361-6528/ab667e/

\item  Y. Tang, S. Zhou, C. Chen, et al., Optics Express, 23 (14), 17923 (2015). DOI:10.1364/OE.23.017923/

\item  Q. Sai, Z. Zhao, C. Xia, et al., Optical Materials, 35, 2155 (2013). DOI:10.1016/j.optmat.2013.05.035/

\item  S. Yamada, M. Yoshimura, S. Sakata, et al., Journal of CityplaceCrystal Growth, 448, 1 (2016). DOI:10.1016/j.jcrysgro.2016.05.003/

\item  Y. Liu, M. Zhang, Y. Nie, et al., Journal of the European Ceramic Society, 37 (15), 4931 (2017). DOI:10.1016/j.jeurceramsoc.2017.06.014/

\item  Y. Nie, J. Han, Y. Liu, et al. Materials Science \& Engineering A, 704, 207 (2017). DOI:10.1016/j.msea.2017.07.098/

\item  U.S. Patent US 2012/0181919 A1 (2012).

\item  EU Patent EP1837921 A1 (2015).

\item  K. Katrunov, V. Ryzhikov, V. Gavrilyuk, S. Naydenov, et al., Nucl. Instrum. and Meth. in Physics Research: Section A, 712, 126 (2013). DOI:10.1016/j.nima.2013.01.065/

\item  Yu. A. Tsirlin, Light Collection in Scintillation Counters, Atomizdat, CityplaceMoscow (1975) [in Russian].

\item  S.V. Naydenov, Technical Physics, 49 (8), 1093 (2004). DOI:10.1134/1.1787678/

\item  V.G. Baryakhtar, V.V. Yanovsky, S.V. Naydenov, A.V. Kurilo, Journal of Experimental and Theoretical Physics, 103 (2), 292 (2006). DOI:10.1134/S1063776106080127/

\item  S.V. Naydenov, Journal of Applied Spectroscopy, 69 (4), 613 (2002). DOI:10.1023/A:1020624720252/

\item  S.V. Nizhankovsky, A.Ya. Dan`ko, V.M. Puzikov, Yu.N. Savvin, et al., Functional materials, 15 (4), 546 (2008). http://dspace.nbuv.gov.ua/handle/123456789/136544/

\item  S.V. Naydenov, B.V. Grinyov, V.D. Ryzhikov, in: IEEE Symposium Conference Record Nuclear Science 2004, Rome, Italy (2004), vol. 2, p. 810. DOI:10.1109/NSSMIC.2004.1462332/

\item  S.V. Nizhankovskyi, A.V. Tan'ko, Y.N. Savvin, et al., Optics and Spectroscopy, 120 (6), 915 (2016). DOI:10.7868/S0030403416050214/
\end{enumerate}

\noindent

\noindent \eject

\noindent \textbf{Figures}

\noindent

\noindent \includegraphics*[width=6.48in, height=3.51in, keepaspectratio=false]{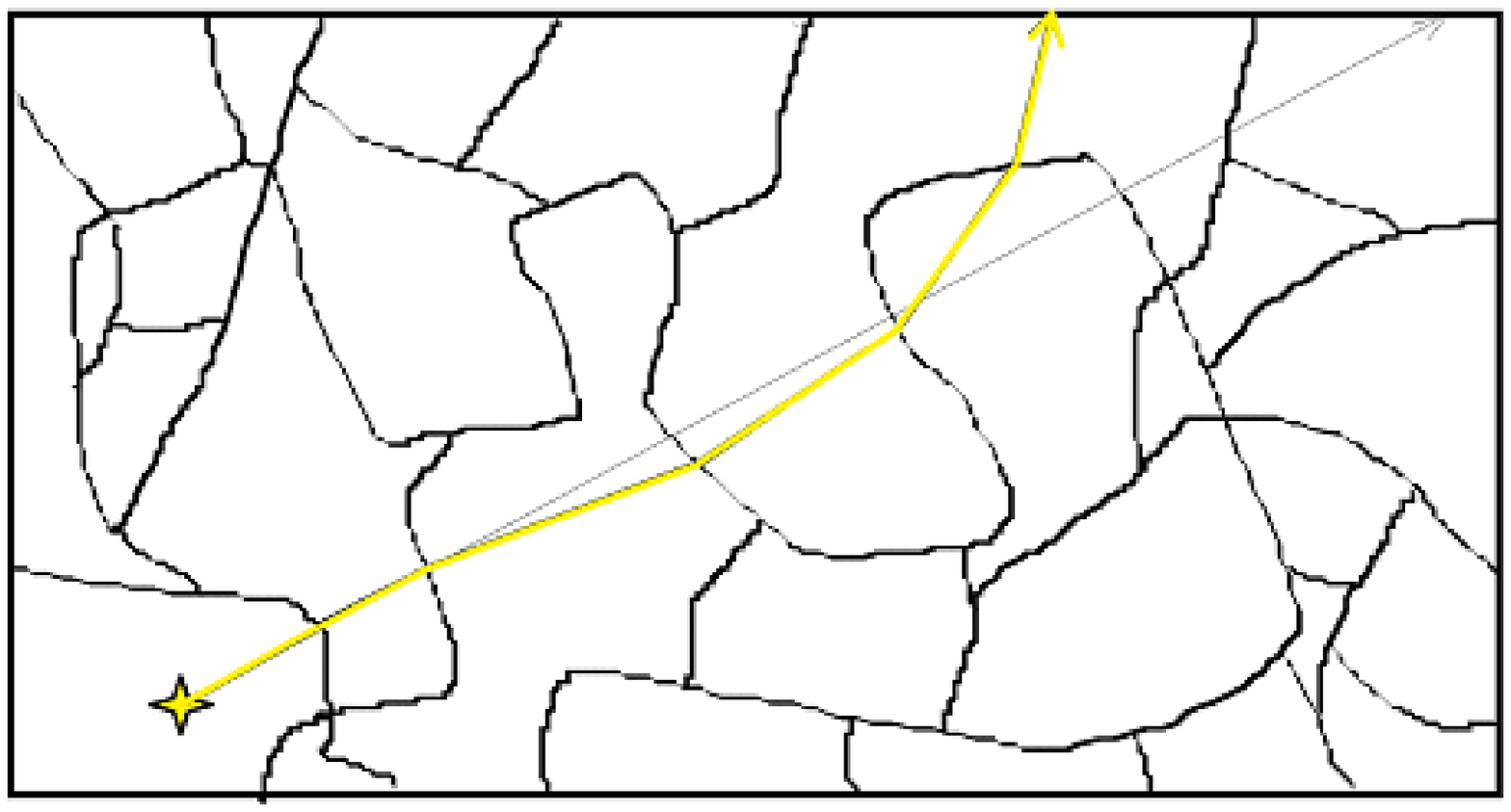}

\noindent 

\noindent \textbf{Fig.~1}. Schematic representation of the propagation of converted light through a eutectic crystal as a heterogeneous light scattering medium. The direction of the beam changes when refracted on (fractal) phase boundaries with different values of the refractive index. The straight line shows the path of the light beam without scattering.

\noindent

\noindent \eject

\noindent

\noindent \includegraphics*[width=5.03in, height=3.26in, keepaspectratio=false]{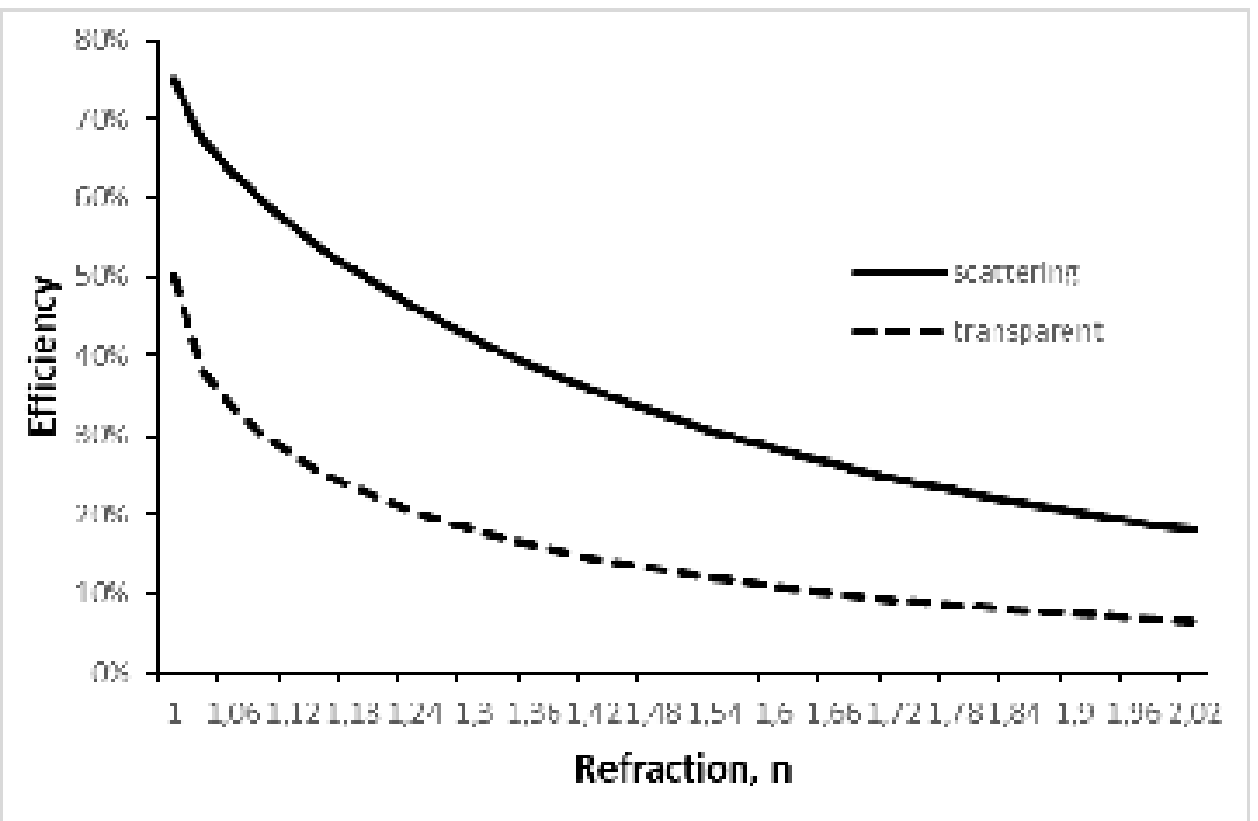}

\noindent 

\noindent \textbf{Fig.~2}. Theoretical dependence of the efficiency of a light converter on the refractive index for a transparent medium with and without scattering. Aperture selected $g=0.5$.

\noindent

\noindent \eject

\noindent

\noindent  \includegraphics*[width=2.08in, height=2.06in, keepaspectratio=false]{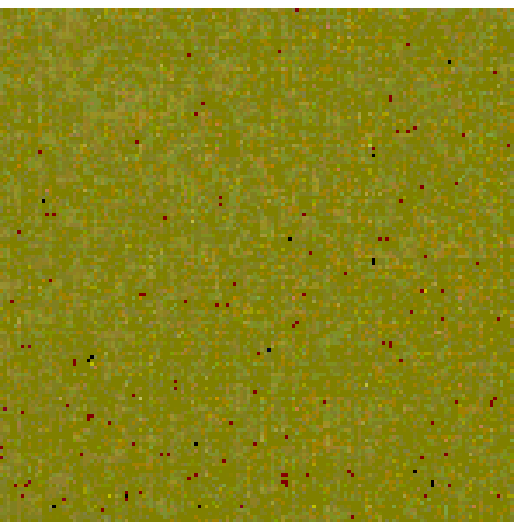} a) \includegraphics*[width=2.72in, height=2.04in, keepaspectratio=false]{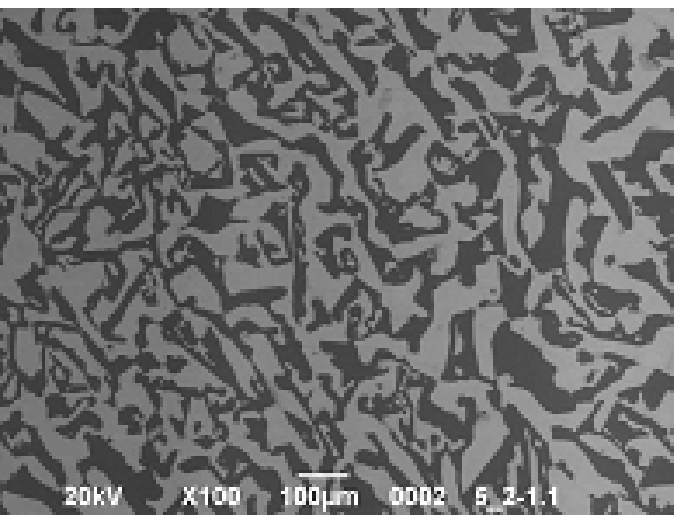} b)

\noindent

\noindent \textbf{Fig.~3}. Crystal of Al$_{2}$O$_{3}$-YAG:Ce eutectic. Photo of a typical specimen with a side length of 3~mm (a); SEM image of the surface fragment (b). Light gray - YAG:Ce phase, dark gray - Al$_{2}$O$_ {3}$ phase.

\noindent

\noindent \eject

\noindent \textbf{Tables}

\noindent

\noindent \textbf{Table~1}. Lighting parameters (absolute and luminous efficiency, color temperature, color coordinates and color rendering index) of converters based on conventional YAG:Ce crystals and Al2O3-YAG:Ce eutectic crystals.

\noindent

\begin{tabular}{|p{0.5in}|p{0.5in}|p{0.5in}|p{0.5in}|p{0.6in}|p{0.6in}|p{0.5in}|p{0.7in}|p{0.5in}|} \hline
\textbf{Sample No.} & \textbf{Crystal Type} & \textbf{Energy Flux, mW } & \textbf{Lumi-nous Flux, lm} & \textbf{Absolute Efficiency, \%} & \textbf{Luminous Efficiency, lm/W} & \textbf{Color Tempe-rature, K} & \textbf{CIE1931 Chroma-ticity Coordinates} & \textbf{Render Index, \%} \\ \hline
1 & Pure\newline YAG:Ce & 0.1932 & 0.06695 & 8.2\% & 28.25 & 6139 & x=0.3099 y=0.4427 & 66.2 \\ \hline
2 & Matte\newline YAG:Ce & 0.3299 & 0.1314 & 13.9\% & 55.44 & 5050 & x=0.3541 y=0.4782 & 63.1 \\ \hline
3 & Eutectic\newline Al${}_{2}$O${}_{3}$-YAG:Ce & 0.3544 & 0.1641 & 15.0\% & 69.24 & 3921 & x=0.4313 y=0.5440 & 41.1 \\ \hline
4 & Eutectic\newline Al${}_{2}$O${}_{3}$-YAG:Ce & 0.3867 & 0.1447 & 16.3\% & 61.05 & 4884 & x=0.3580 y=0.4415 & 67.2 \\ \hline
5 & Eutectic\newline Al${}_{2}$O${}_{3}$-YAG:Ce & 0.3816 & 0.1685 & 16.1\% & 71.10 & 3722 & x=0.4416 y=0.5316 & 49.0 \\ \hline
\end{tabular}

\noindent

\end{document}